# Molecular engineering of metal-free perovskite MDABCO-NH₄I₃ towards enhanced ferroelectric polarization

Xiaoming Wang*, Yanfa Yan*

Department of Physics and Astronomy, and Wright Center for Photovoltaics Innovation and Commercialization, The University of Toledo, Toledo, Ohio 43606, US.

* xiaoming.wang@utoledo.edu, yanfa.yan@utoledo.edu

**ABSTRACT:** Molecular ferroelectrics have attracted increasing interests over the past decade due to mechanical flexibility, chemical diversity, environmental friendliness, easy-processing, and lightness. The performance of molecular ferroelectrics is approaching more and more competitive to their inorganic counterparts. Despite the chemical diversity and tunability of the molecular cations or anions, molecular ferroelectrics are not abundant. In particular, physical directed design principles for new discovery or performance optimization are still lacking. Here, through first-principles calculations, we firstly reveal the importance of the molecular dipole moment of the polar cations in a molecular perovskite ferroelectric, and then propose a molecular dipole guided design rule for high-performance molecular ferroelectrics. Finally, the rule is validated by first-principles calculations. We anticipate that the rule is very useful in the field urging for new and high-performance molecular ferroelectrics.

## I. INTRODUCTION

Ferroelectricity, property that exhibits switchable spontaneous polarization ($P_s$) by the application of external electrical fields, is widely used in capacitors, sensors, actuators, transducers, and nonvolatile memories, etc.[1] Although inorganic perovskite oxides, e.g., $BaTiO_3$ (BTO) and $Pb(Zr,Ti)O_3$ (PZT), with their high performance still dominate the applications after a century of research[2], molecular ferroelectrics are attracting increasing interests due to the advantages as mechanical flexibility, chemical diversity, environmental friendliness, easy-processing, and lightness.[3–6] Past decade has seen great advances of molecular ferroelectrics, e.g., comparable performance in terms of $P_s$[7–10] and piezoelectric coefficient[11] has been achieved compared to their conventional inorganic counterparts.

Despite the chemical diversity and tunability of molecules, molecular ferroelectrics are not abundant. Electric polarization is defined as the dipole moment density, **P** = **p**/V with **p** being the dipole moment and V being the volume. A natural way to make molecular ferroelectrics is to crystalize polar molecules to polar structures.[12,13] Thiourea was the first example of this type,[14] which shows order-disorder like ferroelectricity. However, early days of explorations for molecular ferroelectrics under such route did not show much success.[3] Then, people moved focus on multi-component molecular systems based on displacive or proton-transfer ferroelectricity, e.g., charge-transfer complexes[3,9,15], supramolecules[3,5], metal organic frameworks[4,16], and hybrid organic-inorganic ferroelectrics[6,11,17]. Over years of research, some targeted design rules based on phenomenological theories from a chemical point of view has been proposed very recently for discovery of new molecular ferroelectrics.[18,19]

Polar molecular ions play an important role in molecular ferroelectrics. It can be introduced to otherwise spherical molecular ions to achieve the lower-symmetry ferroelectric phase as in the quasi-spherical theory[18,19], facilitate the ferroelectricity by enhancing the dipole moment as in the H/F substitution approach[19], or even behave as dipole carriers[17]. Physical understanding or first-principles calculations of the dipolar contributions has never been reported. Although new molecular ferroelectrics are continuing to be discovered with the proposed phenomenological rules, physical directed design principle is still lacking. In this work, we take the first step towards designing molecular ferroelectrics from a physical point of view. In particular, we find that the dipolar contribution of the molecular ions to the total polarization is so important that a linear relationship between the molecular dipole moment and $P_s$ is observed through first-principles density-functional theory (DFT) calculations. This leads to the molecular dipole guided design principle for enhancing the performance of molecular ferroelectrics. We take a recent synthesized metal-free three-dimensional (3D) perovskite ferroelectric, namely, MDABCO-NH₄I₃ (MDABCO: N-methyl-N'-diazabicyclo[2.2.2]octonium)[7] as an example to show the importance of the molecular dipoles.

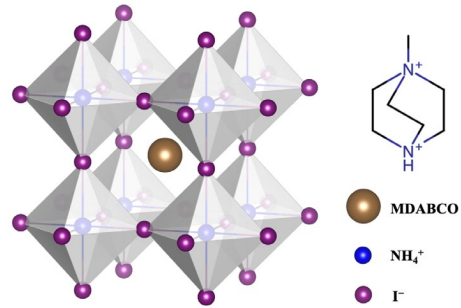

Figure 1. Perovskite structure of MDABCO-NH₄I₃. The iodine anion I⁻, molecular cations MDABCO and NH₄⁺ are denoted by purple, brown and blue balls, respectively. Top right shows the chemical structure of MDABCO.

## II. COMPUTATIONAL DETAILS

First-principles DFT calculations were performed using the Vienna ab initio Simulation Package[20,21] with projector augmented-wave[22] potentials. The exchange-correlation interactions were treated with the generalized gradient approximation of the Perdew-Burke-Ernzerhof (PBE) parametrization[23]. The wave function was expanded in a plane-wave basis set with an energy cutoff of 600 eV. The Brillouin zone was sampled on a Γ centered 4×4×4 mesh. The lattice parameters and atomic positions were fully relaxed until stress and forces were <0.05 GPa and 0.01 eV/Å, respectively. We also relaxed the lattice with

Grimme's D3[24] correction to include the van der Waals interactions. In consistent with previous calculations[25], we found that PBE is better for the current materials. The difference of the polarizations is within 5%. The polarization branch was automatically constructed using the pymatgen package[26,27]. The Wannier function centers of the valence electrons were calculated with the wannier90 code[28].

III. RESULTS AND DISCUSSIONS

A. Polarization of MDABCO-NH$_4$I$_3$

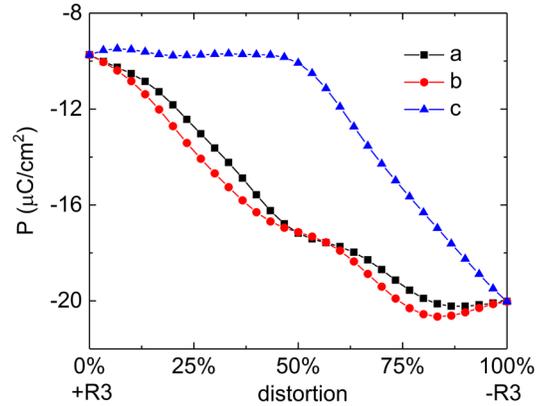

Figure 2. Ferroelectric polarizations along the distortion path from +R3 to −R3 structures. Squares, circles, and triangles denote the crystallographic a, b, and c axes, respectively.

MDABCO-NH$_4$I$_3$ crystalizes in ABY$_3$ (Y = I, Br, Cl) perovskite structure (Figure 1) with B site NH$_4^+$ occupying the center of the octahedra composed of iodine atoms. The A site MDABCO cation (+2) fills the cavities surrounded by the octahedra. MDABCO belongs to the point group 3(C$_3$) with the C$_3$ axis along the [111] direction of the R3 lattice. Direct inspection of MDABCO reveals that a permanent dipole moment along the C$_3$ axis pointed from the methyl group to the globular DABCO exists. The dipole moment is aligned with P$_s$ which is measured to be 22 μC/cm$^2$.[7] Hence, it is expected that the dipole moment of MDABCO contributes to P$_s$. However, previous Bohn charge analysis[25] reveals that the dipolar contribution is negligible based on the small distance between positive and negative charge centers. We note that this conclusion is not accurate because even for small distance large amount of charges can still result in significant dipole moments. Since theoretical calculations of P$_s$ have never been reported, we firstly perform first-principles DFT calculations with the Berry phase method in the framework of modern theory of polarization (MTP)[29–34] to evaluate the P$_s$ of MDABCO-NH$_4$I$_3$. Within MTP, a reference phase, usually the paraelectric phase, is used to calculate the polarization difference from the ferroelectric phase, which leads to P$_s$. We take the inverted structure, namely, the structure with opposite polarization, as the reference phase because the paraelectric phase belongs to space group P432, where MDABCO is completely disordered[7]. The inverted structure, denoted as −R3, can be obtained by applying inversion operation on the coordinates of the ferroelectric structure. The transformation path from +R3 to −R3 is achieved by continuously rotating both MDABCO and NH$_4^+$ along the [111] direction by 180°, whereas the distortions of the iodine atoms are purely displacive. The polarization branches along crystallographic **a**, **b**, and **c** axes are shown in Figure 2. The smooth and much smaller variation of the polarization, compared with the polarization quantum which is calculated to be 55.7 μC/cm$^2$, along the distortion path confirms that the calculated polarizations are in the same branch. The calculated P$_s$ as half the difference of the polarizations between +R and −R structures is 9.6 μC/cm$^2$. The discrepancy with experiments is unclear now. Possible reasons could be interface or surface effects during the measurement that are not taken into account in our calculations. Similar deviations are also found in some other molecular ferroelectrics.[15,35]

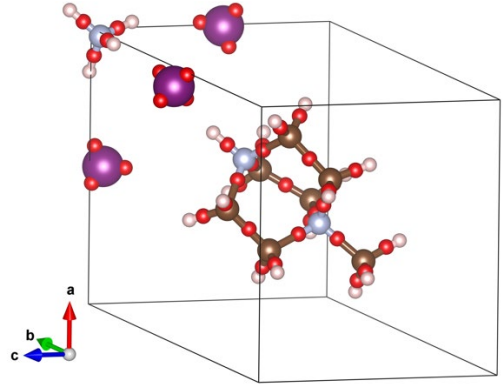

Figure 3. Wannier centers (red balls) in MDABCO-NH$_4$I$_3$. Purple, brown, gray, and pink balls denote the iodine, carbon, nitrogen and hydrogen atoms, respectively.

B. Wannier function analysis

To analyze the contributions to the P$_s$ from different components, we employ maximally-localized Wannier function (WF)[36] which is complementary to the Berry phase method in MTP. In Wannier representation, Bloch electrons can be treated as point charges located at Wannier centers, as shown in Figure 3. We divide MDABCO-NH$_4$I$_3$ into three units: MDABCO, NH$_4$, and I$_3$. It is clear that the Wannier centers can be unambiguously grouped into each unit. For each unit, we define average charge center as $\mathbf{R} = 1/Q \sum_i q_i \mathbf{r}_i$, where $i$ runs over all the ionic positions and Wannier centers with $\mathbf{r}_i$ the corresponding coordinates and $q_i$ the ionic charge or −2 for each WF, $Q$ is the total charge. The charges of MDABCO, NH$_4$, and I$_3$ are +2, +1, and −3, respectively, which equal exactly to their formal charges. The direct coordinates of each unit are tabulated in Table 1, where only one number is shown because the R3 symmetry requires the coordinates along three crystallographic axes to be equal. With this simplified picture, it is easy to infer the coordinates of each unit in the paraelectric phase. The calculated displacements are also shown in Table 1. The displacement of NH$_4$ ($d_B$) relative to I$_3$ is 0.75 Å, which is equivalent to the displacement relative to the center of the [I$_6$] octahedron. Similarly, the displacement of MDABCO ($d_A$) relative to center of the six surrounding octahedra is 0.82 Å. Spontaneous polarization can be simply calculated as P$_s$ = $(2d_A + d_B)/V$ = 9.6 μC/cm$^2$. The contributions from the off-centering-displacement (OCD) of MDABCO and NH$_4$ are calculated to be 6.6 μC/cm$^2$ and 3.0 μC/cm$^2$, respectively. We note in passing that Wannier analysis is more accurate than Bohn charge analysis[25] because P$_s$ can be exactly reproduced in Wannier representations.

**Table 1. Direct coordinates of the average charge centers of MDABCO, NH$_4$, I$_3$ in ferroelectric and paraelectric phases and the corresponding displacements.**

| unit | ferroelectric | paraelectric | displacement (Å) |
|---|---|---|---|
| MDABCO | 0.530 | 0.500 | 0.415 |
| NH$_4$ | 1.025 | 1.000 | 0.345 |
| I$_3$ | 0.471 | 0.500 | -0.406 |

The displacements in Table 1 are calculated based on the center of charge (COC) which is to be different from the center of mass (COM) as usually used in inorganic ferroelectrics or experiments. The COM displacement of MDABCO is 0.137 Å, much smaller than that of COC. The difference is due to the dipole carried by MDABCO. The dipole moment can be calculated as $\mathbf{p} = Q\mathbf{R}$. However, $\mathbf{p}$ is ill-defined and origin-dependent for charged molecules. With COC as origin, $\mathbf{p}$ is always $\mathbf{0}$, which is unphysical. With COM as origin, as we'll show in the next section, counterintuitive $\mathbf{p}$ dependence results. We propose to use the center of the gravity of atomic core electrons, or equivalently the center of ionic charges (COIC), as the origin. With this choice, the calculated dipole moments of MDABCO and NH$_4^+$ are 2.9 D and 0.2 D, respectively, with the corresponding polarizations of 2.4 µC/cm$^2$ and 0.1 µC/cm$^2$. From the charge center analysis, the distance between the centers of positive and negative charges of MDABCO is 0.011 Å, slightly larger than that from Born charge analysis[25]. MDABCO carries a positive charge, sum of the core electrons, of +54, and a negative charge, total number of WFs times −2, of −52. It is natural to expand the charge distribution by monopole, which is located at positive charge center, and dipole of 52 $e$ × 0.011 Å = 2.9 D. The polarization from monopole of MDABCO is purely displacive and amounts to 4.2 µC/cm$^2$. To sum up, we find that the ferroelectricity of MDABCO-NH$_4$I$_3$ is A site dominated with both displacive and order-disorder contributions. In particular, the dipole moment of MDABCO contributes as high as 25% of P$_s$.

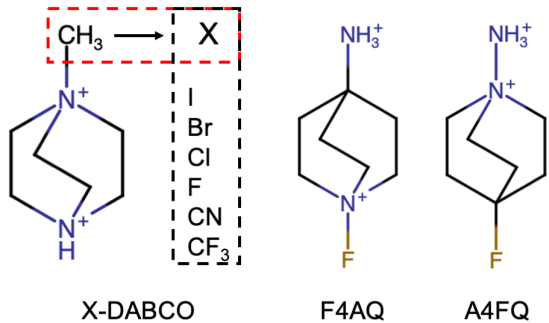

Figure 4. Proposed A cations to enhance the dipole moment of MDABCO. F4AQ: 1-fluoro-4-ammonio-quinuclidium; A4FQ: 1-ammonio-4-fluoro-quinuclidium.

C. Molecular design

Although the dipolar contribution to P$_s$ is not dominant, tuning the dipole moment of molecular cations is, however, simple and easy, which directly leads to the molecular dipole guided design rule. The dipole of MDABCO forms by the introduction of the methyl group to the symmetric DABCO cation, which has zero dipole moment. Indeed, there is no ferroelectricity observed for DABCO-NH$_4$I$_3$[7]. Since dipole is pointed from the methyl group to DABCO, replacing the methyl group with some strong electron-withdrawing groups (EWGs) is expected to enhance the dipole moment of A cation, thus resulting in improved polarization. With this in mind, we replace the methyl group (CH$_3$) with halogens (I, Br, Cl, F), trifluoromethyl group (CF$_3$), and cyano group (CN). The substituted A cations (Figure 4) are denoted as X-DABCO. The calculated dipole moments for isolated A cations with COIC as origin are shown in Table 2. For halogens, the dipole moment shows similar trend as electronegativity, i.e., F > Cl > Br > I. Note that choosing COM as origin, the above order is completely reversed (Table S1), which is counterintuitive. The dipole moment does not necessarily follow the trend of the strength of the inductive effect. For example, CH$_3$ is less electron-withdrawing than I but MDABCO has a larger dipole moment due to different charge distributions. Another way to enhance the dipole moment is to increase the distance between positive and negative centers. We add fluorine and ammonium to the 1 and 4 sites of quinuclidium to form 1-fluoro-4-ammonio-quinuclidium (F4AQ), which turns out to show larger dipole moment (Table 2) than F-DABCO due to the larger distance of opposite charge centers. A step further based on F4AQ, we exchange the N atom of 1 site and C atoms of 4 site to make 1-ammonio-4-fluoro-quinuclidium (A4FQ), which, as expected, shows very large dipole moment.

**Table 2. Dipole moments (in unit of D) of A cations. X-DABCO denote cations that with X replaced by EWGs.**

| X-DABCO | | | | | | | F4AQ | A4FQ |
|---|---|---|---|---|---|---|---|---|
| CH$_3$ | I | Br | Cl | F | CN | CF$_3$ | | |
| 1.7 | 1.4 | 2.3 | 3.1 | 3.8 | 6.2 | 9.1 | 7.2 | 12.4 |

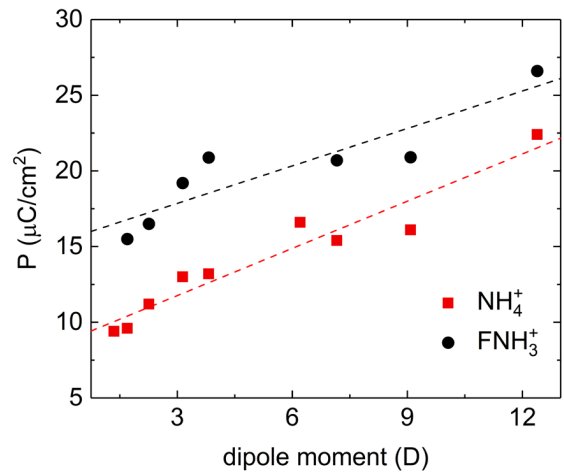

Figure 5. Polarization of ABI$_3$ as a function of the dipole moment of A cation. The red squares and black circles denote NH$_4^+$ and FNH$_3^+$, respectively, as B cation. Linear fittings are shown by dash lines.

We calculate the P$_s$ of A-NH$_4$I$_3$ with different A cations shown in Figure 4. The results are shown by red squares in Figure 5. In general, a linear relationship between the dipole moment of A cation and the spontaneous polarization is observed. The nonmonotonic variation is due to structural relaxations, i.e., A-NH$_4$I$_3$ with different A cations show different ion displacements which contribute dominantly to P$_s$. The increase of the dipole moment of A cation increases the effective OCD, resulting in increased A contributions as expected. Interestingly, the increase of OCD of A cation also slightly increases the OCD of B cation due to Coulombic repulsion (Figure S1). The combined effect leads to consistent increase of the polarization. Same principle can be applied to B cation. For example, replacing NH$_4^+$ with FNH$_3^+$, the dipole moment of B cation increases from 0.2 D to 5.1 D. The calculated P$_s$ of MDABCO-FNH$_3$I$_3$ is increased to 15.5 μC/cm$^2$. The P$_s$ of ABI$_3$ with FNH$_3^+$ as B cation are shown by black circles in Figure 5. As can be seen, the polarization of ABI$_3$ with B cation of FNH$_3^+$ is consistently larger than that of NH$_4^+$. Further improvement of the polarization of ABI$_3$ can be achieved by replacing I with more electronegative Br or Cl, since the polarization can be considered as from the dipole moments between A and B cations and halogen anions. The calculated P$_s$ are summarized in Table 3. For all the studied ABY$_3$ structures, the P$_s$ shows ABCl$_3$ > ABBr$_3$ > ABI$_3$. For example, the P$_s$ of MDABCO-NH$_4$Br$_3$ and MDABCO-NH$_4$Cl$_3$ are 11.6 μC/cm$^2$ and 13.5 μC/cm$^2$, respectively. The improvement of the polarization from engineering the A cation, B cation, and halogen anion are additive. In combination, the P$_s$ of 3.6 times enhancement is achieved compared to the original MDABCO-NH$_4$I$_3$.

Table 3. Polarizations (μC/cm$^2$) of ABY$_3$ with A of X-DABCO, F4AQ, A4FQ, B of NH$_4^+$, FNH$_3^+$, Y of I, Br, Cl. The slash means that the structure of that composition breaks during relaxation.

| A \ Y | X-DABCO CH$_3$ | I | Br | Cl | F | CN | CF$_3$ | F4AQ | A4FQ |
|---|---|---|---|---|---|---|---|---|---|
| **NH$_4^+$** | | | | | | | | | |
| I | 9.6 | 9.4 | 11.2 | 13.0 | 13.2 | 16.6 | 16.1 | 15.4 | 22.4 |
| Br | 11.6 | 13.0 | 14.8 | 16.0 | 15.5 | 20.8 | 19.3 | 18.8 | 26.3 |
| Cl | 13.5 | 16.0 | 17.9 | 19.0 | 18.6 | 24.9 | 22.4 | 24.1 | 30.0 |
| **FNH$_3^+$** | | | | | | | | | |
| I | 15.5 | / | 16.5 | 19.2 | 20.9 | / | 20.9 | 20.7 | 26.6 |
| Br | 18.1 | / | 20.5 | 22.6 | 24.5 | / | 24.0 | 24.4 | 31.1 |
| Cl | 20.6 | / | 24.0 | 25.7 | 28.3 | / | 26.4 | 28.0 | 34.9 |

Finally, we note that in all the calculations 3D perovskite structures with R3 symmetry are assumed for ABY$_3$. Many factors may impact the stability of the structures, which in practice, one may refer to some phenomenological rules[7,19]. We tabulated the Goldschmidt's tolerance factors in Table S2, from which we can estimate the stability from one of the many perspectives.

## IV. CONCLUSION

In conclusion, we calculate the spontaneous polarization of metal-free 3D perovskite MDABCO-NH$_4$I$_3$ through first-principles DFT calculations. Wannier function analysis reveals that the dipole moment of the A cation contributes to as much as 25% of the total polarization. Based on the analysis, a molecular dipole guided design rule towards high-performance ferroelectrics is proposed and validated by DFT calculations. The design rule is expected to apply for any molecular ferroelectrics that have polar cations or anions with order-disorder like behaviors.

## ACKNOWLEDGMENT
The research was funded by the National Science Foundation under contract number DMR-1807818. The method development was supported by Hybrid Organic Inorganic Semiconductors for Energy (CHOISE), an Energy Frontier Research Center funded by the DOE Office of Basic Energy Sciences, Office of Science through the U.S. Department of Energy under Contract No. DE-AC36-08GO28308 with Alliance for Sustainable Energy, Limited Liability Company (LLC), the Manager and Operator of the National Renewable Energy Laboratory. The calculations used resources of the National Energy Research Scientific Computing Center (NERSC), a U.S. Department of Energy Office of Science User Facility located at Lawrence Berkeley National Laboratory, operated under Contract No. DE- AC02-05CH11231, and resources sponsored by the Department of Energy's Office of Energy Efficiency and Renewable Energy and located at the National Renewable Energy Laboratory.

# Supporting information

# Molecular engineering of metal-free perovskite MDABCO-NH$_4$I$_3$ towards enhanced ferroelectric polarization

Xiaoming Wang[*], Yanfa Yan[*]

Department of Physics and Astronomy, and Wright Center for Photovoltaics Innovation and Commercialization, The University of Toledo, Toledo, Ohio 43606, US.

**Supplementary Figures**

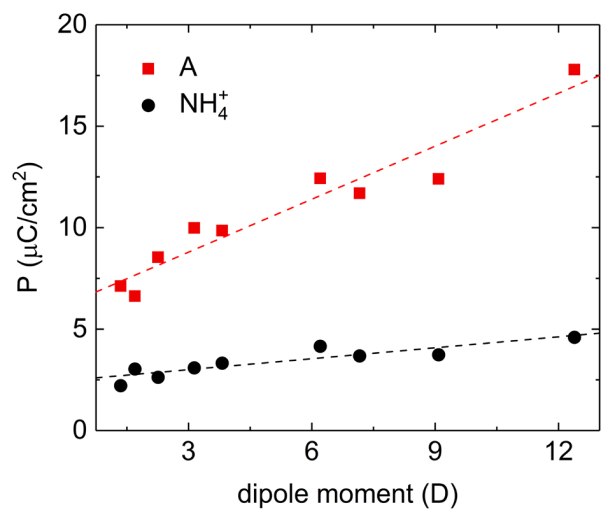

Figure S1. Polarization of A-NH$_4$I$_3$ contributed from OCDs of A and NH$_4^+$ as a function of dipole moment of A cation.

**Supplementary Tables**

**Table S1. Dipole moments (D) of X-DABCO with X being halogens calculated based on different origins.**

|      | I    | Br   | Cl  | F   |
|------|------|------|-----|-----|
| COIC | 1.4  | 2.3  | 3.1 | 3.8 |
| COM  | 13.4 | 10.3 | 6.0 | 4.3 |

**Table S2. Goldschmidt's tolerance factors of ABY$_3$ with A of X-DABCO, F4AQ, A4FQ, B of NH$_4^+$, FNH$_3^+$, Y of I, Br, Cl. The Shannon ionic radii of halogens (Cl$^-$: 1.8 Å, Br$^-$: 2.0 Å, I$^-$: 2.2 Å) are used. For NH$_4^+$ and FNH$_3^+$, values of 1.5 Å and 1.8 Å are adopted. For A cations, we use the furthest interatomic distance.**

| A \ Y | X-DABCO | | | | | | | F4AQ | A4FQ |
|---|---|---|---|---|---|---|---|---|---|
|  | CH$_3$ | I | Br | Cl | F | CN | CF$_3$ | | |
| NH$_4^+$ | | | | | | | | | |
| I  | 0.95 | 0.98 | 0.96 | 0.94 | 0.90 | 0.97 | 1.01 | 0.99 | 0.99 |
| Br | 0.97 | 1.00 | 0.98 | 0.95 | 0.91 | 0.99 | 1.03 | 1.01 | 1.01 |
| Cl | 0.98 | 1.01 | 0.99 | 0.96 | 0.92 | 1.00 | 1.05 | 1.02 | 1.02 |
| FNH$_3^+$ | | | | | | | | | |
| I  | 0.88 | /    | 0.88 | 0.86 | 0.83 | /    | 0.93 | 0.91 | 0.91 |
| Br | 0.89 | /    | 0.89 | 0.87 | 0.84 | /    | 0.94 | 0.92 | 0.92 |
| Cl | 0.90 | /    | 0.90 | 0.88 | 0.84 | /    | 0.95 | 0.93 | 0.93 |